\documentclass[conference]{IEEEtran}
\IEEEoverridecommandlockouts
\usepackage{cite}
\usepackage{amsmath,amsthm, amssymb,amsfonts,latexsym}
\usepackage{algorithmic}
\usepackage{graphicx}
\usepackage{subcaption}
\usepackage{textcomp}
\usepackage{xcolor}
\def\BibTeX{{\rm B\kern-.05em{\sc i\kern-.025em b}\kern-.08em
    T\kern-.1667em\lower.7ex\hbox{E}\kern-.125emX}}
\usepackage{mathtools}
\usepackage{comment}
\usepackage{hyperref}
\usepackage{booktabs} % for professional tables
\usepackage{url}            % simple URL typesetting
\usepackage{xcolor}

\ifCLASSINFOpdf
\else
\fi

\usepackage{amssymb}
\usepackage{amsmath}
\usepackage{titlesec}
\usepackage{url}
\usepackage{makecell}
\usepackage{xcolor}

\begin{document}
\title{Scalable and Interpretable Quantum Mixture of Experts for High-Dimensional Spatio-Temporal Data}
% \thanks{The views expressed in this article are those of the authors and do not represent the views of Wells Fargo. This article is for informational purposes only. Nothing contained in this article should be construed as investment advice. Wells Fargo makes no express or implied warranties and expressly disclaims all legal, tax, and accounting implications related to this article.}}

% Concise form
% \author{\IEEEauthorblockN{Anonymous Authors}}
\author{Junghoon Justin Park$^1$, Anna Jung$^2$, Kyumin Lee$^3$, Samuel Yen-Chi Chen$^4$, \\ Huan-Hsin Tseng$^5$  Shinjae Yoo$^5$, Jiook Cha$^{1,6,7}$\\
\small $^1$Interdisciplinary Program in Artificial Intelligence, Seoul National University \\ $^2$Department of English Language and Literature, Seoul National University \\
$^3$School of Civil, Architectural Engineering and Landscape Architecture, Sungkyunkwan University \\ $^4$Wells Fargo $^5$Artificial Intelligence Department, Brookhaven National Laboratory\\$^6$Department of Brain and Cognitive Sciences, Seoul National University $^7$Department of Psychology, Seoul National University \\
\small \texttt{utopie9090@snu.ac.kr, annajung227@snu.ac.kr, supermini713@skku.edu,} \\
\small \texttt{ycchen1989@ieee.org, htseng@bnl.gov, sjyoo@bnl.gov, connectome@snu.ac.kr}
}


\maketitle

\begin{abstract}
Quantum Mixture of Experts (QMoE) architectures offer a promising path toward scalable quantum machine learning, yet existing approaches remain largely confined to low-dimensional benchmarks and often rely on na\"ive input partitioning that disrupts feature dependencies. We address these gaps by introducing Feature-Partitioned QMoE, which partitions high-dimensional spatio-temporal inputs into functionally coherent feature groups and assigns each group to a dedicated quantum expert. We further propose a hierarchical interpretability framework that leverages the modular structure of QMoE to produce circuit-level gating analysis, per-expert gradient saliency, and heterogeneity-based subtype discovery. Applied to ADHD classification from resting-state fMRI ($N=4{,}458$, ABCD Study, 180 brain ROIs, 363 timepoints), the proposed Theory-Driven QMoE achieves performance comparable to matched-capacity classical and quantum baselines while enabling circuit-attributable interpretation. The framework identifies FDR-corrected ADHD-associated biomarkers in executive-control and anterior default-mode networks and reveals three neurobiologically meaningful ADHD subtypes. These results position Feature-Partitioned QMoE as a scalable and interpretable methodology for applying quantum machine learning to high-dimensional biomedical data.
\end{abstract}


% \fi
\IEEEpeerreviewmaketitle

% \section{Introduction}}
\section{Introduction}

Quantum Machine Learning (QML) leverages superposition and entanglement to model complex problems in high-dimensional Hilbert spaces, with applications spanning chemistry, finance, high-energy physics, healthcare, and materials science \cite{Biamonte,Cerezo2022,Gujju,Peral,Wen,McArdle,Herman,Egger,Chen_HEP,DiMeglio,Mensa,Ullah,Cong}. However, practical deployment remains constrained by NISQ hardware limitations, including limited qubit counts, short coherence times, and sparse qubit connectivity \cite{Cerezo2022}.

Multi-chip ensemble variational quantum circuits (VQCs) have been proposed to improve scalability and generalization by distributing computation across multiple quantum circuits, achieving strong performance on image classification, EEG analysis, and reinforcement learning tasks \cite{MultiChip,MultiChip_SuperMario}. Nevertheless, existing approaches scale computational cost with the number of experts and typically partition features using simple index-based splits that ignore intrinsic feature dependencies, leading to information loss in structured data such as spatio-temporal signals \cite{MultiChip}.

Quantum Mixture of Experts (QMoE) addresses these limitations by combining sparse expert routing with specialized quantum experts, activating only the most relevant circuits for each input while better preserving complex feature relationships \cite{QMoE,Jacobs,shazeer2017}. This provides a scalable framework for modeling high-dimensional datasets such as resting-state fMRI.

The main contributions are summarized as follows:
(1) We propose \textbf{Feature-Partitioned QMoE}, a quantum mixture-of-experts architecture demonstrated on high-dimensional biomedical time series with 180 ROIs, 363 timepoints, and $N=4{,}458$ subjects;
(2) We introduce a \textbf{circuit-aligned feature partitioning strategy} that assigns each quantum expert to a neuroscience-defined functional circuit;
(3) We show that theory-driven partitioning preserves substantially more functional connections within experts (\textbf{62.4\%} vs.\ \textbf{43.9\%}) while maintaining comparable predictive performance;
(4) We develop an \textbf{architecture-agnostic hierarchical interpretability framework} based on gating analysis, per-expert saliency, and heterogeneity-based subtype discovery.

\section{Background}
\subsection{Variational Quantum Circuit}
A Variational Quantum Circuit (VQC) maps a classical input vector $\vec{x}$ into a high-dimensional Hilbert space via an encoding circuit $U(\vec{x})$, producing the state $U(\vec{x})|0\rangle^{\otimes n}$. This state is transformed by a parameterized ansatz $W(\Theta)$, defined as a product of $L$ trainable layers:
\begin{equation}
    |\psi(\vec{x},\Theta)\rangle = W(\Theta)U(\vec{x})|0\rangle^{\otimes n} = \left(\prod_{j=L}^{1} V_j(\vec{\theta}_j)\right) U(\vec{x})|0\rangle^{\otimes n}
    \label{eq:vqc}
\end{equation}
Information is extracted via the expectation values of Hermitian observables $\hat{H}_k$:
\begin{equation}
    \langle \hat{H}_k \rangle = \langle \psi(\vec{x},\Theta) | \hat{H}_k | \psi(\vec{x},\Theta) \rangle
\end{equation}
The VQC implements a parametric function $\vec{f}(\vec{x};\Theta) = (\langle \hat{H}_1 \rangle, \dots, \langle \hat{H}_n \rangle)$, which is optimized by minimizing a loss function $\mathcal{L}(\vec{x}, y; \Theta)$---such as the mean squared error $\sum_i \|\vec{f}(\vec{x}_i;\Theta) - \vec{y}_i\|^2$---through a hybrid quantum-classical loop.

\subsection{Quantum Time-series Transformer}

The Quantum Time-series Transformer (QTSTransformer) replaces the quadratic complexity of classical self-attention with quantum primitives based on the linear combination of unitaries (LCU) and quantum singular value transformation (QSVT) \cite{QTSTransformer}. Sequence embeddings are represented as a weighted superposition,
\begin{equation}
    M = \sum_j b_j U_j,
\end{equation}
where $U_j$ denotes a quantum-encoded sequence segment. QSVT then applies a polynomial transformation
\begin{equation}
    P_{\vec{c}}(M)=c_dM^d+\cdots+c_1M+c_0I,
\end{equation}
enabling nonlinear processing analogous to classical attention mechanisms \cite{Childs,Gilyen}. This approach achieves $\mathcal{O}(\mathrm{polylog}\,T)$ complexity for sequence length $T$, making it attractive for long-sequence modeling \cite{QTSTransformer}.

\subsection{Quantum Mixture of Experts}

Recent work has extended the Mixture of Experts (MoE) paradigm to quantum machine learning to improve model capacity and scalability. Proposed approaches include native QMoE architectures with quantum routing \cite{QMoE}, mixtures of trainable quantum unitaries \cite{Coyle}, infinite-width Mixtures of Quantum Experts (MoQE) \cite{Tognini}, multimodal quantum MoE frameworks \cite{Dutta}, and hybrid classical--quantum expert ensembles \cite{chaves2026}.

\section{Feature-Partitioned Quantum Mixture of Experts}
\label{sec:quantum_moe}
\begin{figure}[t]
\centering
\includegraphics[width=\columnwidth]{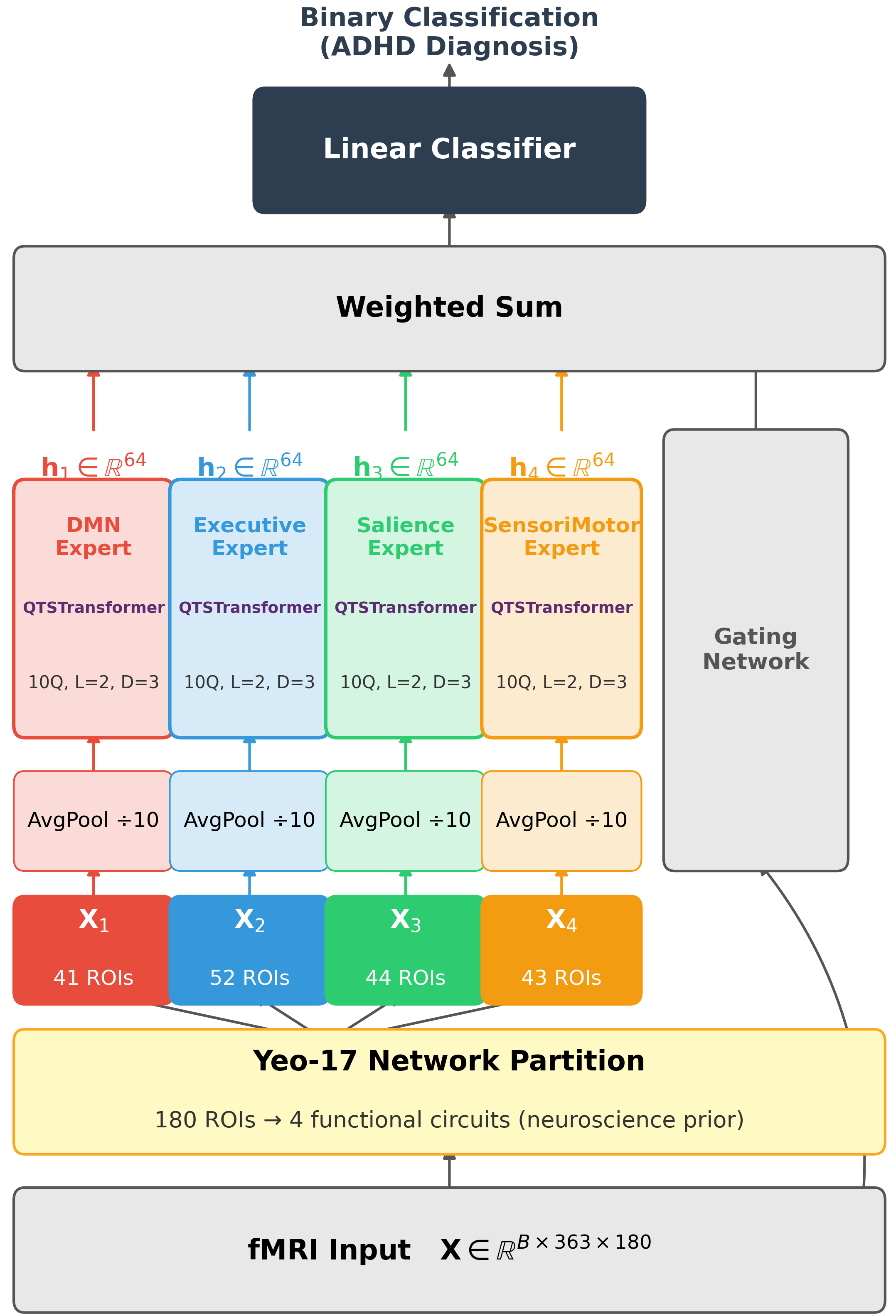}
\caption{Architecture overview of the Theory-Driven QMoE. Theory-Driven QMoE partitions 180 fMRI ROIs into four neuroscience-validated functional circuits (DMN, Executive, Salience, and SensoriMotor) based on the Yeo 17-network atlas.}
\label{fig:architecture}
\end{figure}

We present a QMoE architecture that partitions input features into functionally coherent clusters, processes each cluster through a dedicated VQC, and combines expert outputs via a learned gating network. The architecture is designed for high-dimensional datasets where input features exhibit structured dependencies that can be exploited through informed partitioning. Figure~\ref{fig:architecture} provides an overview of the proposed theory-driven partitioning strategy.

\subsection{Feature Partitioning}
\label{sec:partitioning}

Consider an input $\mathbf{X} \in \mathbb{R}^{T \times C}$ with $T$ timesteps and $C$ features. We partition the features into $K$ disjoint subsets $\mathcal{S}_1, \ldots, \mathcal{S}_K$ with $\bigcup_k \mathcal{S}_k = \{1,\ldots,C\}$, yielding per-expert inputs $\mathbf{X}_k = \mathbf{X}[:, \mathcal{S}_k] \in \mathbb{R}^{T \times |\mathcal{S}_k|}$.

\subsubsection{Theory-Driven Partitioning}

When domain knowledge provides a validated grouping of features into functionally coherent categories, this knowledge directly defines the partition via a mapping $\mathcal{G}: \{1,\ldots,C\} \to \{1,\ldots,K\}$:
\begin{equation}
    \mathcal{S}_k = \{ c \mid \mathcal{G}(c) = k \}
    \label{eq:theory_partition}
\end{equation}
Examples include mapping spatial regions to known functional systems, grouping sensor channels by modality, or assigning variables by established causal structure. The key assumption is that features within each cluster share dependencies relevant to the prediction task.

\subsection{Quantum Expert Architecture}
\label{sec:quantum_expert}

Each expert $k$ is a hybrid classical-quantum model that maps $\mathbf{X}_k$ to a fixed-dimensional output $\mathbf{h}_k \in \mathbb{R}^{H}$. A classical encoder maps the input to rotation angles, a VQC applies parameterized single-qubit rotations and two-qubit entangling gates to an $n_q$-qubit register, and observable expectation values are mapped to $\mathbf{h}_k$ via a classical output layer:
\begin{align}
    \mathbf{h}_k &= \mathbf{W}_{\mathrm{out}} \left[\langle \psi_k | O_j | \psi_k \rangle \right]_{j=1}^{N_O} + \mathbf{b}_{\mathrm{out}}, \nonumber \\
    |\psi_k\rangle &= W(\Theta_k)\,U(\mathcal{E}_k(\mathbf{X}_k))\,|0\rangle^{\otimes n_q}
    \label{eq:expert}
\end{align}
where $\mathcal{E}_k$ is the classical encoder, $U(\cdot)$ is the data-encoding unitary, $W(\Theta_k)$ is the trainable ansatz parameterized by $\Theta_k$ (cf.\ Eq.~\eqref{eq:vqc}), and $\{O_j\}_{j=1}^{N_O}$ are measurement observables (e.g., single-qubit Pauli operators). The framework is agnostic to the specific ansatz. In the feature-partitioned setting, the entangling gates within the VQC are particularly relevant because they couple features that share genuine dependencies within each cluster rather than creating spurious correlations across unrelated features. For spatio-temporal inputs, temporal integration can be performed via strategies such as pooling, sequential processing, or a linear combination of unitaries~\cite{Childs}. The specific ansatz and temporal strategy used in our experiments are described in the Experiments section~\ref{sec:experiments}.

\subsection{Gating Network and Prediction}
\label{sec:gating}

A two-layer gating network with Gaussian exploration noise~\cite{shazeer2017} computes soft routing weights from the temporal mean of the input:
\begin{align}
    \mathbf{g} &= \mathrm{softmax}\!\left(\mathbf{W}_2 \, \mathrm{ReLU}(\mathbf{W}_1 \bar{\mathbf{x}} + \mathbf{b}_1) + \mathbf{b}_2 + \boldsymbol{\epsilon}\right) \in \mathbb{R}^{K}, \nonumber \\
    \bar{\mathbf{x}} &= \tfrac{1}{T}\textstyle\sum_{t=1}^{T} \mathbf{X}^{(t)}.
    \label{eq:gating}
\end{align}
The final prediction is $\hat{y}=h(\sum_k g_k\mathbf{h}_k)$, where $h$ is the classifier head. Training minimizes the task loss with a load-balancing regularizer~\cite{shazeer2017,Fedus2022},
$\mathcal{L}=\mathcal{L}_{\mathrm{task}}+\alpha\mathcal{L}_{\mathrm{balance}}$,
which encourages uniform expert utilization.

\subsection{Hierarchical Interpretability Framework}
\label{sec:interpretability}

Because each expert processes only its assigned feature subset, the architecture enables interpretability at three nested levels that are native to the partitioned model. A monolithic single-expert model can only approximate these analyses by post-hoc grouping of its input gradient, without physical expert isolation; the feature-partitioned architecture instead yields per-expert, confound-free attributions by construction.

\subsubsection{Level 1: Cluster-Level Gating Analysis}

The gating weights $\mathbf{g} \in \mathbb{R}^K$ provide a per-sample routing profile. Comparing mean gate weights between classes via statistical testing (e.g., Welch's $t$-test) identifies which feature clusters are differentially engaged:
\begin{equation}
    \Delta g_k = \mathbb{E}[g_k \mid y\!=\!1] - \mathbb{E}[g_k \mid y\!=\!0]
    \label{eq:gate_diff}
\end{equation}

\subsubsection{Level 2: Per-Expert Gradient Saliency}

Per-expert saliency is obtained by reducing the expert Jacobian $\partial \mathbf{h}_k / \partial \mathbf{X}_k \in \mathbb{R}^{H \times T \times |\mathcal{S}_k|}$ to a per-feature vector via averaging over the hidden ($H$) and time ($T$) axes:
\begin{equation}
    \mathbf{s}_k[c] = \frac{1}{HT}\sum_{h=1}^{H}\sum_{t=1}^{T} \frac{\partial h_{k,h}}{\partial X_{k,t,c}} \in \mathbb{R}^{|\mathcal{S}_k|}, \quad c \in \mathcal{S}_k.
    \label{eq:saliency}
\end{equation}
This isolates what each expert learned within its cluster. Unlike model-level saliency ($\partial \hat{y} / \partial \mathbf{X}$), which conflates all experts and routing, per-expert saliency is computed with respect to expert $k$'s input only, providing two complementary views: \emph{absolute} saliency $|\mathbf{s}_k|$ (magnitude of influence) and \emph{signed} saliency $\mathbf{s}_k$ (direction toward positive or negative class). Aggregating over known feature sub-groupings within each cluster reveals \emph{intra-cluster feature hierarchies}.

\subsubsection{Level 3: Heterogeneity Analysis via Multi-Level Clustering}

The architecture provides three nested representation spaces for data-driven subtype discovery within a class of interest:
\begin{align}
    \text{Level 1:} \; & \mathbf{g}^{(i)} \in \mathbb{R}^{K} & \text{(routing profile)} \nonumber \\
    \text{Level 2:} \; & [\mathbf{h}_1^{(i)}, \ldots, \mathbf{h}_K^{(i)}] \in \mathbb{R}^{KH} & \text{(expert representations)} \nonumber \\
    \text{Level 3:} \; & \mathbf{s}^{(i)} = [\mathbf{s}_1^{(i)};\ldots;\mathbf{s}_K^{(i)}] \in \mathbb{R}^{C} & \text{(feature saliency)} \nonumber
\end{align}
where the Level-3 vector is the concatenation of the per-expert saliencies in Eq.~\eqref{eq:saliency} (since $\sum_k |\mathcal{S}_k| = C$). Subjects are clustered at each level and characterized by their dominant profiles. Cross-level concordance, measured via the Adjusted Rand Index (ARI)~\cite{Hubert1985, Steinley}, assesses whether subtypes are consistent across representational scales. This yields subtypes defined by how the model \emph{processes} each subject through its learned, task-optimized representations, rather than by post-hoc analysis of raw data. Crucially, the domain interpretability of these subtypes depends on the partitioning strategy: theory-driven partitioning produces cluster-attributed, domain-interpretable subtypes.

\section{Experiments}
\label{sec:experiments}

\subsection{Dataset and Rationale}

We evaluate Feature-Partitioned QMoE on resting-state fMRI from the Adolescent Brain Cognitive Development (ABCD) Study~\cite{Casey} for binary ADHD classification.

Resting-state fMRI provides a high-dimensional spatio-temporal benchmark with structured functional dependencies between brain regions, making it well suited for evaluating feature partitioning. ADHD also has well-established functional circuits~\cite{Castellanos2012,Cortese2012,Nigg2020,Koirala2024}, enabling a neuroscience-informed partition into the Default Mode, Executive Control, Salience, and SensoriMotor networks. Finally, the large feature dimension ($C=180$ ROIs) relative to the capacity of individual quantum circuits ($n_q=10$ qubits) naturally motivates a multi-expert architecture.

\subsection{Data Preprocessing}

The ABCD resting-state fMRI data was parcellated using the HCP-MMP1 atlas~\cite{Glasser}. Following prior work~\cite{QTSTransformer, MBBN}, bilateral homologous regions were averaged to obtain $C = 180$ cortical ROIs. Each scan consisted of $T = 363$ timepoints (TR $= 0.8$\,s). ADHD diagnosis was defined using the Child Behavior Checklist, with attention and ADHD T-scores $> 65$ classified as ADHD-positive; healthy controls had no diagnosed mental health disorders~\cite{Cordova}. Subjects with fewer than 300 timepoints were excluded, and shorter scans were zero-padded to 363 timepoints. Per-ROI z-score normalization was computed from the training set and values were clipped at $\pm10$ standard deviations. The final dataset comprised $N = 4{,}458$ subjects, split into 70\%/15\%/15\% training/validation/test sets using stratified random sampling ($N_{\mathrm{train}} = 3{,}120$, $N_{\mathrm{val}} = 669$, $N_{\mathrm{test}} = 669$).

\subsection{Quantum Expert Implementation}

Each quantum expert is a QTSTransformer~\cite{QTSTransformer}. Given its assigned ROI subset $\mathbf{X}_k \in \mathbb{R}^{363 \times |\mathcal{S}_k|}$, temporal average pooling ($p=10$) reduces the sequence to $\tilde{T}=36$ timesteps, followed by sinusoidal positional encoding~\cite{Vaswani}. A learned linear projection with sigmoid activation maps the features to rotation angles,
$\boldsymbol{\theta}^{(t)} = 2\pi \cdot \sigma(\mathbf{W}\tilde{\mathbf{X}}_k^{(t)}+\mathbf{b})$.

The variational quantum circuit uses $n_q=10$ qubits and $L=2$ ansatz layers (Eq.~\eqref{eq:vqc}) with alternating $RY$ rotations and $CRX$ entangling gates. Temporal integration is performed using a classically simulated linear combination of unitaries~\cite{Childs}, followed by a degree-$D=3$ polynomial transformation inspired by quantum singular value transformation~\cite{Gilyen}. Pauli-$X/Y/Z$ measurements produce a $3n_q=30$-dimensional representation, which is projected to the expert embedding $\mathbf{h}_k \in \mathbb{R}^{64}$.

\subsection{QMoE Model Configurations}
\label{sec:qmoe_configs}

We train a Theory-Driven QMoE with four experts ($K=4$) over 180 brain ROIs. The ROIs are partitioned into four neuroscience-validated functional circuits based on the Yeo 17-network atlas~\cite{Yeo2011} and prior ADHD literature~\cite{Castellanos2012,Cortese2012}:
(1) \textbf{Default Mode Network (DMN)}: 41 ROIs;
(2) \textbf{Executive Control}: 52 ROIs
(3) \textbf{Salience}: 44 ROIs
(4) \textbf{SensoriMotor}: 43 ROIs.
This theory-driven partition preserves known functional organization and provides interpretable expert assignments.

\subsection{Baseline Models}
\label{sec:baselines}

We compare against representative classical and quantum baselines trained using the same protocol (learning rate $10^{-3}$, batch size 32, cosine annealing, and early stopping with patience 20):

\begin{itemize}
    \item \textbf{Classical Single Expert}: A Transformer processing all 180 ROIs jointly (134,849 parameters).
    \item \textbf{Classical 4-Expert MoE}: Four classical Transformer experts with theory-driven ROI partitioning and learned gating (516,485 parameters).
    \item \textbf{Quantum Single Expert}: A single QTSTransformer processing all 180 ROIs (16,936 parameters).
    \item \textbf{Quantum Arbitrary-Split MoE}: Four QTSTransformer experts with arbitrary ROI partitions matched only in size to the theory-driven split (34,885 parameters).
\end{itemize}

The proposed Theory-Driven QMoE contains 34,885 trainable parameters, matching the arbitrary-split baseline and closely matched by the compact classical MoE (31,917 parameters); this parameter-matched control lets us compare quantum and classical experts at comparable capacity rather than against the much larger conventional MoE (516,485 parameters). All models were implemented in PyTorch 2.5.0 with PennyLane 0.42.3~\cite{Bergholm2018} and trained on NVIDIA A100 (80\,GB HBM) GPUs on the NERSC Perlmutter supercomputer.

\section{Results}
\label{sec:results}

% \subsection{Classification Performance}

Table~\ref{tab:results} summarizes the performance of all models on the ABCD ADHD binary classification task.

\begin{table*}[t]
\centering
\caption{ADHD classification results on the ABCD dataset ($N=4{,}458$).}
\label{tab:results}
\small
\begin{tabular}{llcc}
\toprule
\textbf{Model} & \textbf{Partitioning} & \textbf{Params} & \textbf{Test AUC} \\
\midrule
Classical SE        & None (all 180 ROIs)    & 134,849 & $0.5997 \pm 0.0174$ \\
Classical MoE       & Theory-driven (Yeo-17) & 516,485 & $0.5686 \pm 0.0246$ \\
Compact Classical MoE & Theory-driven (Yeo-17) & 31,917 & $0.5873 \pm 0.0025$ \\
\midrule
Quantum Arb MoE     & Arbitrary (contiguous) & 34,885 & $0.5576 \pm 0.0151$ \\
\textbf{Theory QMoE} (ours) & \textbf{Theory-driven (Yeo-17)} & 34,885 & $\mathbf{0.5631 \pm 0.0177}$ \\
\bottomrule
\end{tabular}
\end{table*}

\subsection{Theory-Driven Partitioning Preserves Functional Dependencies}

We compare two QMoE models with identical architecture and parameter count (34,885 parameters) but different ROI assignments: theory-driven Yeo-17 partitioning and arbitrary index-based splitting. Both achieve comparable test AUC ($0.5631 \pm 0.0177$ vs.\ $0.5576 \pm 0.0151$; paired $t$-test across seeds, $p=0.61$), indicating that the neuroscience-informed partition incurs no accuracy penalty while enabling circuit-level interpretation.

On the group-average training functional-connectivity matrix, the theory-driven partition retains \textbf{62.4\%} of the top-5\% strongest functional connections within experts, compared with \textbf{43.9\%} for arbitrary splitting (chance: 24.8\%; identical expert sizes). This demonstrates that theory-driven partitioning better preserves functional dependencies while supporting hierarchical interpretability.

\subsection{Hierarchical Interpretability}

We apply the three-level interpretability framework (Section~\ref{sec:interpretability}) to the Theory-Driven QMoE (10 qubits, degree~3, 4 experts). We focus on this model because its experts process neuroscience-defined circuits, making per-expert saliency directly circuit-attributable. Figure~\ref{fig:interpretability} presents four key visualizations.

\subsubsection{Cluster-Level Gating Analysis}

Gate weights remain nearly uniform across the four experts (${\sim}0.250 \pm 0.003$), with no significant differences between ADHD and control subjects (all $p>0.27$). This reflects the load-balancing regularization ($\alpha=0.1$), indicating that classification relies on expert specialization rather than class-dependent routing.

\subsubsection{Per-Expert Gradient Saliency}

Per-expert saliency ($\partial \mathbf{h}_k / \partial \mathbf{X}_k$) is evaluated at the network level via Welch's $t$-test between ADHD and control subjects, with Benjamini--Hochberg false-discovery-rate (FDR) correction applied within each expert's per-ROI family. Four network-level associations survive correction ($q<0.05$), all led by the Executive expert: ContC ($q=0.0004$, positively associated with ADHD), DorsAttnB ($q=0.003$, positively associated), and ContA ($q=0.032$, negatively associated), together with the anterior default-mode sub-network DefaultA ($q=0.027$, positively associated). Two individual ROIs survive ROI-level correction: FST ($q=0.007$, +ADHD) and POS2 ($q=0.027$, $-$ADHD). Area~8C, the top-saliency ROI within the FDR-surviving ContC network, is elevated in ADHD and also emerges in the classical 4-expert model, but does not itself survive ROI-level correction; the robust signal is carried at the network level by ContC.

Figure~\ref{fig:interpretability}a shows distinct intra-circuit hierarchies. The DMN expert prioritizes the anterior sub-network DefaultA over the posterior DefaultC, consistent with the anterior--posterior DMN dissociation in ADHD~\cite{Castellanos2012}. The Salience expert identifies Limbic\_OFC and Limbic\_TempPole as negatively associated with ADHD, whereas SalVentAttnA/B are positively associated; these Salience associations are nominal and do not survive FDR correction.

Figure~\ref{fig:interpretability}b shows clear expert specialization, with each expert assigning the highest weights to its designated Yeo-17 subnetworks, supporting the effectiveness of the theory-driven partition.

\subsubsection{Heterogeneity Analysis (ADHD Subtype Discovery)}

\begin{figure*}[htp]
\centering
\begin{tabular}{cc}
\includegraphics[width=0.48\textwidth]{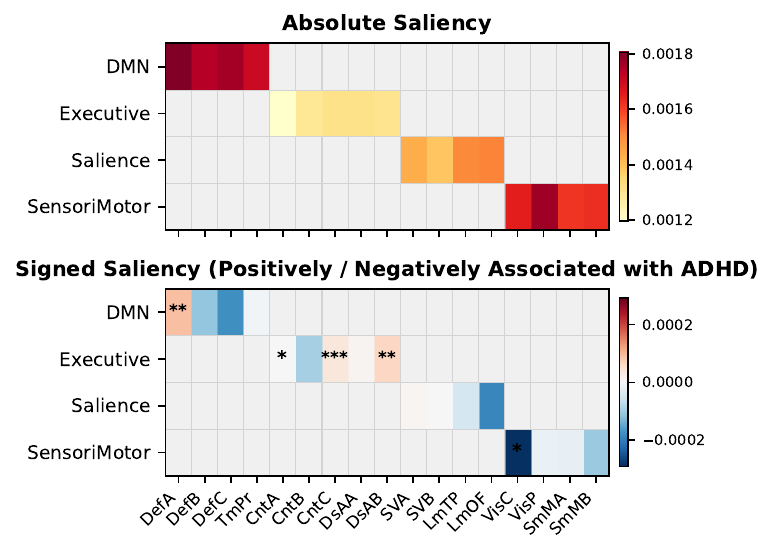} &
\includegraphics[width=0.48\textwidth]{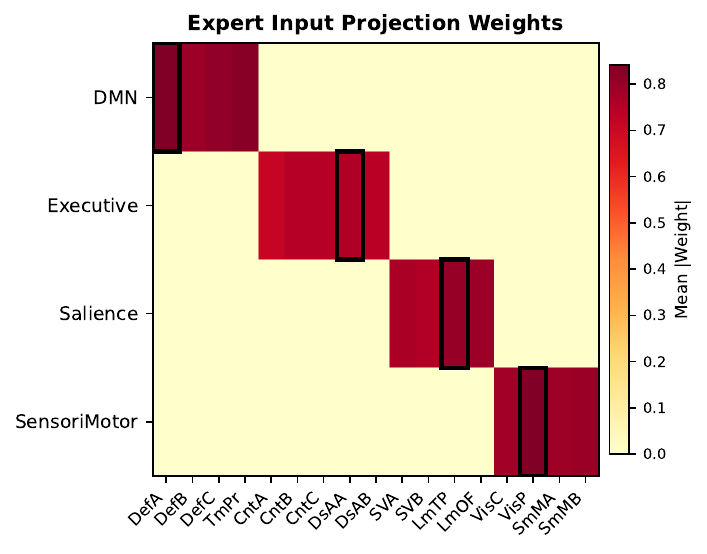} \\
\small (a) Per-expert network saliency & \small (b) Expert input projection weights \\[6pt]
\includegraphics[width=0.48\textwidth]{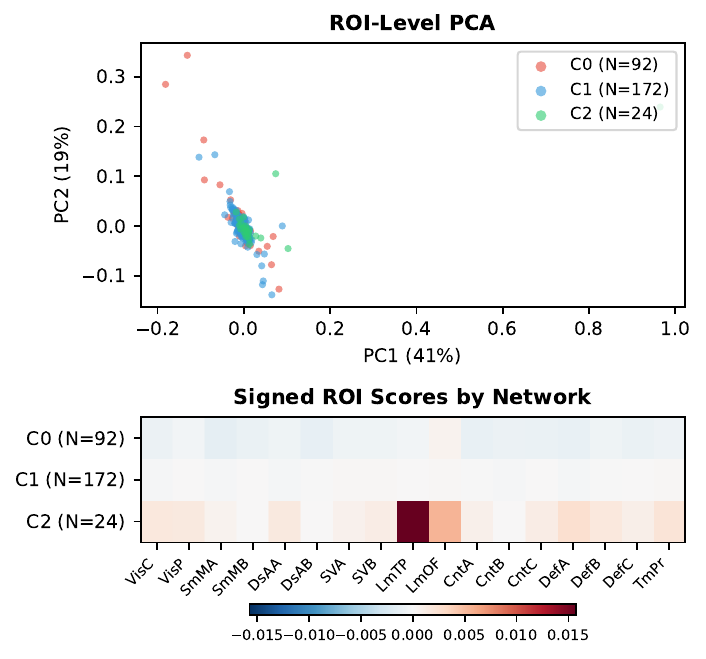} &
\includegraphics[width=0.42\textwidth]{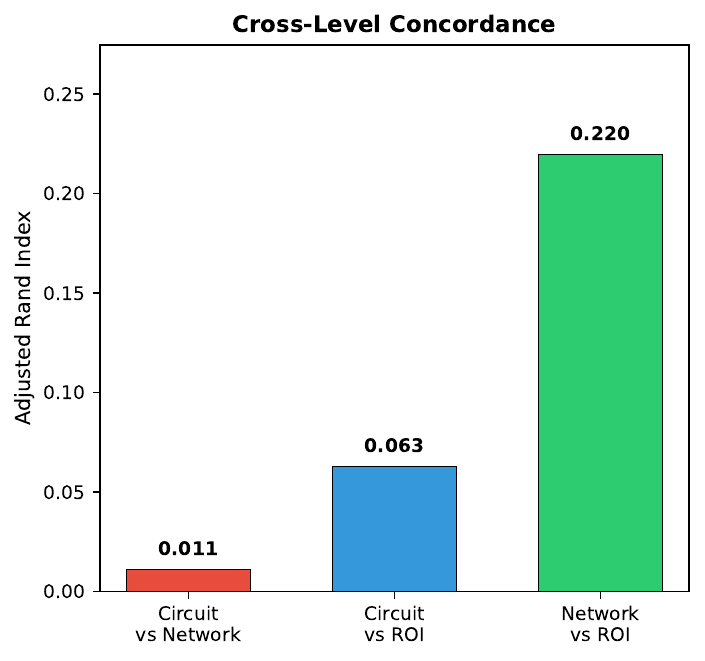} \\
\small (c) ROI-level heterogeneity clusters & \small (d) Cross-level concordance \\
\end{tabular}
\caption{Hierarchical interpretability of the Theory-Driven QMoE (10 qubits, 4 experts). Red/blue denote positive/negative association with ADHD.
\textbf{(a)} Per-expert network saliency. Top: absolute saliency; bottom: signed saliency with Welch's $t$-test markers ($*p<0.05$, $**p<0.01$, $***p<0.001$; nominal/uncorrected---the FDR-surviving associations are those reported in the text).
\textbf{(b)} Expert input projection weights grouped by Yeo-17 sub-networks; black borders indicate the highest-weighted network for each expert.
\textbf{(c)} ROI-level clustering ($K=3$) of ADHD subjects. Cluster~2 ($N=24$, 8.3\%) exhibits a prominent limbic temporal pole signature and is also identified by the classical 4-expert model.
\textbf{(d)} Cross-level Adjusted Rand Index (ARI) between Circuit (gate weights), Network (expert outputs), and ROI (saliency) representations; Network--ROI ARI $=0.220$ indicates partial subtype consistency across scales.}
\label{fig:interpretability}
\end{figure*}

Multi-level clustering of subjects with ADHD ($N=288$) identifies three neurobiological subtypes at the ROI level (Figure~\ref{fig:interpretability}c). The largest subtype (Cluster~1, 59.7\%) is characterized by elevated SalVentAttnA activation, the second (Cluster~0, 31.9\%) by increased Limbic\_OFC importance, and a rare subtype (Cluster~2, 8.3\%, $N=24$) by strong Limbic\_TempPole involvement, suggesting distinct attentional, reward-processing, and memory/emotional dysfunction profiles, respectively. The rare subtype is also recovered by the classical 4-expert model, indicating that it reflects genuine neurobiological structure rather than a model artifact.

Cross-level concordance (Figure~\ref{fig:interpretability}d) shows very low Circuit--Network agreement (ARI = 0.011), consistent with near-uniform gate weights, but moderate Network--ROI agreement (ARI = 0.220), indicating that ADHD heterogeneity is primarily captured at the network and ROI levels rather than by coarse circuit routing.

\section{Conclusion}

We presented Feature-Partitioned QMoE, a quantum mixture-of-experts architecture that addresses three limitations of existing multi-chip quantum ensembles: naïve feature partitioning, limited scalability to high-dimensional data, and the lack of systematic interpretability.

More broadly, Feature-Partitioned QMoE provides a scalable and interpretable framework for high-dimensional scientific data. Although demonstrated using neuroscience-defined functional circuits, the proposed feature-partitioning strategy naturally extends to other structured domains with meaningful feature groupings, such as gene regulatory networks and multimodal sensor data. Future work will investigate data-driven expert partitioning when domain-specific feature groupings are unavailable. The proposed interpretability framework—combining gating analysis, per-expert saliency, and heterogeneity clustering—generalizes to any MoE architecture with disjoint feature subsets, providing a systematic approach for extracting scientific insights from quantum models.

\section*{\textbf{Acknowledgment}}
{\small
The authors extend their gratitude to the members of the Connectome Lab for their invaluable support and critical contributions to the development of this study. This work was supported by the National Research Foundation of Korea (NRF) grant funded by the Korea government (MSIT) (No. 2021R1C1C1006503, RS-2023-00266787, RS-2023-00265406, RS-2024-00421268, RS-2024-00342301, RS-2024-00435727, RS-2025-25457239, RS-2021-NR061370, NRF-2021M3E5D2A01022515, and NRF-2021S1A3A2A02090597), by the Researchers Program through Seoul National University (No. 200-20250071, 200-20250049, 200-20250116, 200-20250115, 200-20250113, 0670-20250039, 200-20260009). Additional support was provided by the Institute of Information \& Communications Technology Planning \& Evaluation (IITP) grant funded by the Korea government (MSIT) [No. RS-2021-II211343 Artificial Intelligence Graduate School Program, Seoul National University] and by the Global Research Support Program in the Digital Field (RS-2024-00421268). This work was also supported by the Artificial Intelligence Industrial Convergence Cluster Development Project funded by the Ministry of Science and ICT and Gwangju Metropolitan City, by the Korea Brain Research Institute (KBRI) basic research program (25-BR-05-01), by the Korea Health Industry Development Institute (KHIDI) and the Ministry of Health and Welfare, Republic of Korea (HR22C1605), and by the Korea Basic Science Institute (National Research Facilities and Equipment Center) grant funded by the Ministry of Education (RS-2024-00435727). We acknowledge the National Supercomputing Center for providing supercomputing resources and technical support (KSC-2023-CRE-0568, KSC-2024-CRE-0198, KSC-2025-CRE-0340). An award for computer time was provided by the U.S. Department of Energy’s (DOE) ASCR Leadership Computing Challenge (ALCC). This research used resources of the National Energy Research Scientific Computing Center (NERSC), a DOE Office of Science User Facility, under ALCC award m4750-2024, and supporting resources at the Argonne and Oak Ridge Leadership Computing Facilities, U.S. DOE Office of Science user facilities at Argonne National Laboratory and Oak Ridge National Laboratory.}

\clearpage

\bibliographystyle{ieeetr}
\bibliography{references}

\end{document}